# The HI Content of Spirals. I. Field-Galaxy HI Mass Functions and HI Mass-Optical Size Regressions

José M. Solanes, Riccardo Giovanelli,

and

Martha P. Haynes
Center for Radiophysics and Space Research and
National Astronomy and Ionosphere Center[1],
Cornell University, Ithaca, NY 14853
e-mail: (solanes, riccardo, haynes)@astrosun.tn.cornell.edu

## ABSTRACT

A standard parametric maximum-likelihood technique is used to determine both the probability distribution over total HI mass $M_{\rm HI}$ and the regression of this quantity on the linear optical diameter $D_{\rm o}$ for field giant spirals (Sa-Sc) from a complete HI-flux-limited data set of these objects. This sample is extracted from a subset of the optical magnitude-limited *Catalog of Galaxies and Clusters of Galaxies* comprised of galaxies observed in the 21 cm HI line and located in the lowest density environments of the Pisces-Perseus supercluster region bounded by $22^{\rm h} \le {\rm R.A.} \le 3^{\rm h}$, $0° \le {\rm Dec.} \le 40°$.

Gaussian and Schechter parametrizations of the HI mass function are explored. We find that the available data are equally well described by both models, and that the different morphological classes of giant spirals have HI mass functions which, in general, agree well within the errors. The largest discrepancy corresponds to the Sb-type systems which exhibit a deficit of low HI-mass objects relative to the other types.

Using a straightforward generalization of the gaussian model, we have also investigated the linear dependence of $M_{\rm HI}$ on $D_{\rm o}$. We confirm that the HI content of spirals is much better predicted by the size of their optical disks than by their morphological types alone. The inferred correlations imply a considerable decrease of the ratio $M_{\rm HI}/D_{\rm o}^2$ with increasing galaxy size for types earlier than Sc.

---

astro-ph/9511003  1 Nov 1995

*Subject headings:* galaxies: evolution — galaxies: fundamental parameters — galaxies: HI mass function — galaxies: ISM — galaxies: spiral — radio lines: galaxies

## 1. Introduction

Because the neutral hydrogen component is both a good indicator of the potential for star formation and an excellent probe of the physical conditions of the galaxy environment, the quantification of HI content is a useful diagnostic tool for the study of galaxy evolution. The knowledge of the probability distribution over total HI mass (the HI mass function, HIMF) for isolated galaxies, and of any dependence it may have on other galaxy properties, play, in particular, a key role in the definition of the standards of HI content used in the investigation of environmental influences on cluster galaxies (Haynes & Giovanelli 1984, HG84). But the HIMF of galaxies has also important cosmological applications that, in many cases, complement those of their optical luminosity function. Thus, the HIMF has been used to place constraints on the space density of HI-rich low-surface-brightness galaxies (Briggs 1990; Hoffman, Lu, & Salpeter 1992), to determine the relative contributions of galaxies of different sizes and morphologies to the current cosmological mass density of neutral gas (Briggs & Rao 1993; Rao & Briggs 1993), and to investigate the existence of large reservoirs of metal-poor gas around high-luminosity spirals (Bothun 1985).

However, unlike the often-studied optical luminosity function, the HIMF is still ill-constrained due to the small size and incompleteness of the HI line data sets. Estimates of the HIMF have been restricted so far to the investigation of its form in specific HI mass ranges, or have been obtained indirectly by reformulating the optical luminosity function in terms of the HI mass, after adopting a power-law description of its dependence on luminosity (see references above). More importantly, the possible biases introduced in the HIMF by the sample selection have not been fully accounted for. For example, nearly all HI line surveys are based on optical catalogs which tend to be biased against the inclusion of low-luminosity/low-surface-brightness galaxies despite the fact that such objects may have high HI masses. Likewise, HI observations include their own selection effects, such as HI-flux sensitivity cutoffs which lead to a distance dependence of the typical HI mass of the targeted objects (Malmquist bias). Finally, HI observations based on very shallow catalogs (i.e., with depths of only a few thousand $km\,s^{-1}$) may: i) introduce large relative errors in the calculation of distance-dependent galaxy properties due to peculiar velocities if redshifts are used to infer distance; and ii) yield HI mass distributions that are not representative of the neutral hydrogen contents of galaxies located at larger distances.

Our main goal in this paper is to determine bias-free HIMFs for the 'field' giant spiral population (types Sa-Sc). In what follows 'field' will always designate regions of low galaxy density in which the HI content is expected to be minimally affected by environmental influences. Our primary HI data set of field objects is presented in the next section, where we describe the selection criteria adopted for its construction, as well as the corrections applied to the raw observational data in the calculation of intrinsic galaxy properties. A complete HI-line-flux-limited subset extracted from the prime field sample is used in Section 3 to construct the HIMF of different morphological subgroups of spirals. The method applied is a standard density-independent maximum-likelihood estimator which takes into account the HI-flux sensitivity limitation of the data. Two different parametrizations of the HIMF are investigated: the gaussian and the Schechter model. In Section 4, the same technique is applied to obtain Malmquist bias-free estimates of the linear dependence of the total HI mass on the linear optical diameter, and the implications of the inferred correlations on the invariance of the global HI surface density for giant spirals are discussed. A summary of the overall results of the paper is presented in Section 5. Throughout the paper we take $H_0 = 100 h \, \mathrm{km \, s^{-1} \, Mpc}$.

## 2. The Data

### 2.1. Sample Selection

Our sample of field galaxies has been extracted from an all-sky private data base maintained by RG and MPH known as the *Arecibo General Catalog* (AGC), which contains an extensive compilation of galaxy parameters gathered from a variety of sources. We have focused on the galaxies listed in the Zwicky et al. (1960-1968) magnitude-limited *Catalog of Galaxies and Clusters of Galaxies* (CGCG) located in the Pisces-Perseus supercluster region bounded by $22^\mathrm{h} \leq \mathrm{R.A.} \leq 3^\mathrm{h}$, $0° \leq \mathrm{Dec.} \leq 40°$, for which the AGC has a wealth of both HI line and optical data (Giovanelli & Haynes 1985a; Giovanelli et al. 1986b; Haynes et al. 1988; Giovanelli & Haynes 1989; Wegner, Haynes, & Giovanelli 1993; Giovanelli & Haynes 1993). In particular, HI line data are available for 65% of all CGCG galaxies within that region and for 90% of spirals with morphological types Sa-Sc.

Since the high-density enhancements in this region outline a filamentary structure which is perpendicular to the line of sight (Giovanelli, Haynes, & Chincarini 1986a), field sample membership has been directly assigned from measures of the local number density of galaxies *projected on the plane of the sky*. For such a structure, this quantity closely mimics the 3-D local density. We have assigned to each CGCG galaxy in the Pisces-Perseus region a galactic-absorption-corrected surface density parameter $\mu$ given by

$$\mu = \frac{5}{\pi s_6^2} \, \mathrm{dex}(0.6 \Delta m_\mathrm{o}^g) \, \mathrm{deg}^{-2} \ . \qquad (1)$$

In equation (1), $s_6$ is the projected distance of the galaxy to its sixth nearest neighbor, in degrees. The adopted number of neighbors allows to retain locality in the estimation of the surface density, while reducing poissonian fluctuations. HI column density measurements and galaxy counts are available for nearly two thirds of the CGCG galaxies within the Pisces-Perseus supercluster region. For these objects the optical ($B$ band) galactic extinction in magnitude units $\Delta m_o^g$ has been calculated using the method of Burstein & Heiles (1978). For the rest, we use a smooth extinction model of the form (Heiles 1976)

$$\Delta m_o^g = 0.25(1/\sin|b^{II}| - 1) . \qquad (2)$$

The adopted field sample includes all galaxies with local surface densities below $3\,\mathrm{deg}^{-2}$. By adopting such an upper limit for $\mu$, we remove from the original data set all the objects in the cores (i.e., with projected distances $\lesssim 1.5$ Abell radius) of the most prominent galaxy clusters that delineate the Pisces-Perseus chain. In such sites, environmental influences on the HI content are most likely to occur (Giovanelli & Haynes 1985b). Only first ranked galaxies of clusters for which the CGCG does not sample a substantial fraction of their luminosity function may appear in regions of low apparent density as measured by $\mu$. However, the bias in the inferred HIMFs caused by this erroneus environment assignment is likely to be minimal because: i) the CGCG is only moderately deep and therefore is expected to contain only a small number of undersampled clusters in the Pisces-Perseus region; and ii) the brightest cluster members are usually of the earliest Hubble types, which will be excluded from the present study (see below). On the other hand, we estimate that less than 10% of the galaxies located in high-apparent-density regions could be actually field objects.

The sky distribution of the CGCG galaxies located within the borders of the Pisces-Perseus region defined above is plotted in Figure 1. The top panel shows the distribution in right ascension and declination of all the CGCG members. Several structures can be identified, the most prominent being the main ridge of the Pisces-Perseus supercluster which extends across the entire range of right ascension dominating the upper declination zones. Note that, because of its location at low galactic latitudes for which the galactic extinction is large, the northernmost section of the ridge containing the Perseus cluster is excluded from this study. In the middle panel, only those galaxies with $\mu \geq 3\,\mathrm{deg}^{-2}$ are shown. As expected, clumpiness is much more conspicuous; the filamentary structure of the Pisces-Perseus supercluster and the galaxy clusters embedded within it are clearly emphasized. Once these overdense regions are removed from the original data set, the sky distribution of the remaining galaxies, plotted in the bottom panel of Figure 1, looks approximately homogeneous.

Figure 2 compares the normalized velocity histograms constructed from all the galaxies with measured redshifts in each one of the subsets depicted in Figure 1, with a smooth curve representing the velocity distribution expected for a spatially homogeneous galaxy set characterized by a universal Schechter optical luminosity function with $M_o^* = -19.2 + 5\log h$ and $\alpha_o = -1.1$ (de Lapparent, Geller, & Huchra 1989). The curves have been corrected for the magnitude incompleteness of the observational sample using the ratio between the number of galaxies with redshift and the total number of CGCG galaxies calculated in bins of 0.1 in apparent magnitude (Giovanelli & Haynes 1982). The theoretical velocity distributions are normalized to have the same area as the observed ones within the selected range of velocities. All velocities are referenced to the rest frame of the Cosmic Microwave Background (CMB; see § 2.2.2).

All three velocity histograms are similar and show a substantial disagreement with respect to the expected distributions in a homogeneous universe. The most important feature in these histograms is the strong confinement of galaxies around $5000\,\mathrm{km\,s^{-1}}$, corresponding to the main ridge of the Pisces-Perseus supercluster (Giovanelli et al. 1986a). Although of lower contrast, the peak near $5000\,\mathrm{km\,s^{-1}}$ is still clearly visible in the velocity distribution for the galaxies in regions of low surface density (bottom panel). The superposed dotted histogram in the bottom panel shows the velocity distribution of galaxies with $\mu$ below $1\,\mathrm{deg}^{-2}$; even in this extreme low density regime, clumping is still evident, implying that the uniformity seen in the sky distribution is only apparent. As the scale of these spatial inhomogeneities is non-negligible compared with the depth of our sample, they will distort the shape of the distribution function of any distance-dependent galaxy property, if the assumption of homogeneity is adopted for its construction (e.g., Davis & Huchra 1982). For instance, a local density enhancement would overestimate the HIMF for galaxies with intrinsically low HI fluxes, which are sampled only nearby. Maximum-likelihood methods, as the one presented in Section 3.1, are density-independent and, hence, insensitive to inhomogeneities in the spatial distribution of galaxies.

Because early-type galaxies are often not detected by HI line observations, galaxies of type earlier than Sa are not included in the present study. Likewise, dwarf irregulars (types Scd or later) are not included because of their high susceptibility to Malmquist and catalog selection biases. We have also excluded all galaxies with recession velocities below $2000\,\mathrm{km\,s^{-1}}$ to minimize distance errors arising from peculiar velocities. This strict minimum velocity cutoff results however in the exclusion of only a few objects due to the scarcity of nearby galaxies in this region caused by the well-known Pisces-Perseus foreground void (e.g., Haynes & Giovanelli 1986; see also Fig. 2). The trimmed-down field sample consists of 934 giant spirals of Hubble types from Sa to Sc. Below, we briefly describe the derivation of the global properties of these objects that are relevant to our investigation.

## 2.2. Galaxy Properties

In this section, we outline the derivation of the intrinsic galaxy properties necessary for performing our analysis of the HI content of field giant spirals. A previous discussion in somewhat more detail is given in HG84.

### 2.2.1. Morphological Type

For galaxies listed in the *Uppsala General Catalog of Galaxies* (Nilson 1973; UGC), we use Nilson's morphological classification. For the remaining objects, the morphological types have been assigned through visual examination of the Palomar Observatory Sky Survey (POSS) prints by RG. No distinction has been made between normal and barred spirals. The typical uncertainty in the morphological classification is $+/-$ one Hubble type, although for some individual galaxies, especially the ones with the smallest angular diameters, the errors may be larger.

### 2.2.2. Radial Distance

Heliocentric radial velocities are taken from the HI line spectrum, except for nondetected field giant spirals for which the optical radial velocity is used. Heliocentric velocities are then converted to the CMB rest frame by correcting them for a solar velocity with respect to that frame of $369.5\,\mathrm{km\,s^{-1}}$ in the direction $(l^{II}, b^{II}) = (264°.4, 48°.4)$ (Kogut et al. 1993). Because of the adopted low velocity cutoff and the nearly anti-Virgo location of the Pisces-Perseus supercluster, no correction for Virgocentric infall is applied. Conversion to radial distance $hr$ is done by assuming euclidean properties of space. Accordingly, the cosmological corrections to the distance-dependent global properties involved in the present study, which are only of the order of a few percent, have also been neglected.

### 2.2.3. Linear Optical Diameter and Inclination

The major and minor angular diameters, $a$ and $b$, are the blue visual sizes listed in the UGC if available. For non-UGC galaxies, diameters have been measured by visual inspection of the POSS blue prints. Apparent diameters are estimated in bins of $0\farcm 1$ for dimensions larger than $1'$ and in bins of $0\farcm 05$ for dimensions below that value. Giovanelli et al. 1994 have shown that visual diameters closely approach face-on isophotal sizes so they do not need to be corrected for the inclination of the parent galaxy (internal absorption). We also neglect the effects of galactic extinction on apparent dimensions. Linear optical diameters are calculated from

$$hD_o = 0.291(hr)a \text{ kpc} . \tag{3}$$

Because in the next sections we will be dealing with this property expressed in a logarithmic scale, we define $\mathcal{D}_o \equiv \log(hD_o)$ to simplify the notation.

The inclinations $i$ with respect to the line of sight are estimated by assuming that spirals are oblate spheroids of intrinsic axial ratio $q$

$$\cos^2 i = \frac{(b/a)^2 - q^2}{1 - q^2} . \tag{4}$$

Intrinsic axial ratios have been taken equal to the modal values of the distributions of true ellipticities of galaxies of different Hubble types in the *Second Reference Catalog of Bright Galaxies* (de Vaucouleurs, de Vaucouleurs, & Corwin 1976) inferred by Binney & de Vaucouleurs (1981). For the giant spirals these are: $q = 0.32$ for Sa; $q = 0.23$ for Sab; and $q = 0.18$ for Sb-Sc. Whenever $b/a < q$, $i$ is set to $90°$.

### 2.2.4. Optical Luminosity

Raw apparent Zwicky magnitudes are corrected for:

*i)* Variations due to the shift in the energy distribution (K-correction). The adopted corrections are based on Pence's (1976) tabulation for the $B$ filter.

*ii)* Systematic variations in the CGCG. Following Kron & Shane (1976) and Giovanelli & Haynes (1984), Zwicky magnitudes are corrected for both their systematic dependence on catalog volume and systematic discrepancies based on magnitude range.

*iii)* Galactic extinction. The same extinction corrections as used in the determination of the surface density parameter (§ 2.1) are applied.

*iv)* Internal absorption. Type-dependent corrections are derived in Appendix A from all the giant spirals in the CGCG.

Optical luminosities are calculated from

$$h^2 L_o = (hr)^2 \text{dex}[10 + 0.4(5.41 - m_o)] L_\odot , \tag{5}$$

where $m_o$ is the blue apparent magnitude corrected for all the above effects and $+5.41$ is the adopted absolute blue magnitude of the Sun (Lang 1974).

*2.2.5. HI Mass and Global HI Surface Density*

Observed HI flux densities are corrected for pointing errors and beam attenuation as described in HG84. Galaxies with HI profiles showing absorption lines or confusion with an obvious optical companion have been discarded. Following the investigation of the HI opacity by Giovanelli et al. (1994) (see also Appendix B in HG84), HI self-absorption corrections to integrated HI fluxes are considered negligible.

The neutral hydrogen mass in a galaxy is calculated from the expression

$$h^2 M_{\rm HI} = 2.36 \times 10^5 (hr)^2 F_{\rm HI} M_\odot \;, \qquad (6)$$

where $F_{\rm HI}$ represents the corrected HI flux density integrated over the profile width in units of Jy km s$^{-1}$. We define $\mathcal{M}_{\rm HI} \equiv \log(h^2 M_{\rm HI})$. For nondetections, an upper limit to HI mass has been estimated assuming that the emission profile is rectangular of amplitude 1.5 times the observed rms noise per channel and width equal to that expected for a galaxy of the same morphological type and luminosity, properly corrected for redshift broadening and viewing inclination.

The global HI surface density is simply defined as

$$\overline{\Sigma}'_{\rm HI} = \log(F_{\rm HI}/a^2) \simeq \mathcal{M}_{\rm HI} - 2\mathcal{D}_{\rm o} - 6.45 \;, \qquad (7)$$

on a logarithmic scale. The prime notation is intended to reflect the fact that this definition of the global HI surface density is a hybrid quantity since the optical diameter is used to indicate area.

## 3. The HIMF

### 3.1. Method

To determine the HIMF for field giant spirals (in what follows HIMF is always meant to designate the *differential* HI mass function), we employ the density-independent parametric maximum-likelihood technique developed by Sandage, Tammann, & Yahil (1979, STY). Maximum-likelihood methods, which are extensively used in the construction of the distribution functions of galaxy properties (e.g., Binggeli, Sandage, & Tammann 1988, and references therein), rely on the hypothesis that the HIMF has a universal form, i.e., that the relative densities of galaxies of different HI masses are everywhere the same. In this case, the normalization of the HIMF is a local measure of density that factors out, allowing the conditional probability $p(\mathcal{M}_{{\rm HI},i}|hr_i)$ that a galaxy at a distance $hr_i$ has the observed HI mass $\mathcal{M}_{\text{HI},i}$ *in a complete HI-flux-limited catalog* to be written simply as

$$p(\mathcal{M}_{\text{HI},i}|hr_i) = \begin{cases} \phi(\mathcal{M}_{\text{HI},i})/\Phi(\mathcal{M}_{\text{HI,lim}}(hr_i)) & \text{if } \mathcal{M}_{\text{HI},i} \geq \mathcal{M}_{\text{HI,lim}}(hr_i) \\ 0 & \text{otherwise} \end{cases}, \qquad (8)$$

where $\phi(\mathcal{M}_{\text{HI}})$ and $\Phi(\mathcal{M}_{\text{HI}})$ are the a priori adopted functional form of the HIMF and its integral, respectively, and $\mathcal{M}_{\text{HI,lim}}(hr)$ is the minimum HI mass detectable at distance $hr$ due to the flux limitation, which is in this manner explicitely taken into account in the construction of the HIMF. Given a parametrization for $\phi(\mathcal{M}_{\text{HI}})$, the free parameters of the model can then be determined by minimizing a computationally convenient likelihood function defined by

$$\Lambda = -2 \sum_i \ln p(\mathcal{M}_{\text{HI},i}|hr_i) . \qquad (9)$$

We have chosen initially a gaussian model for the HIMF

$$\phi(\mathcal{M}_{\text{HI}})d\mathcal{M}_{\text{HI}} \propto \exp\{-0.5[(\mathcal{M}_{\text{HI}} - \langle\mathcal{M}_{\text{HI}}\rangle)/\sigma_{\mathcal{M}_{\text{HI}}}]^2\}d\mathcal{M}_{\text{HI}} , \qquad (10)$$

for this is the functionality predicted by the gaussianity of the optical luminosity function of giant spirals (Binggeli et al. 1988) and the rough direct proportionality between $\mathcal{M}_{\text{HI}}$ and $\log(h^2 L_o)$. Nonetheless, in order to test how well the available data constrains the shape of the HIMF, we have also tried a Schechter function

$$\phi(\mathcal{M}_{\text{HI}})d\mathcal{M}_{\text{HI}} \propto [\text{dex}(\mathcal{M}_{\text{HI}} - \mathcal{M}^*_{\text{HI}})]^{\alpha_{\text{HI}}+1} \exp[-\text{dex}(\mathcal{M}_{\text{HI}} - \mathcal{M}^*_{\text{HI}})]d\mathcal{M}_{\text{HI}} . \qquad (11)$$

Because the probability function defined by equation (8) is independent of the local space density of galaxies, the STY method gives an estimate of the HIMF which is not biased by inhomogeneities in the galaxy spatial distribution. Other interesting features of this technique are its suitability for small samples since the binning of the data is avoided, and the possibility of obtaining accurate confidence limits for the adjustable parameters from their error contours

$$\Lambda = \Lambda_{\min} + \Delta\chi^2(\nu, \beta) , \qquad (12)$$

where $\Delta\chi^2(\nu, \beta)$ is the value of a $\chi^2$ distribution with $\nu$ degrees of freedom (free parameters in the minimization) and significance $\beta$. In addition, the likelihood of the assumed HIMF model, which is not assessed by the STY method, can be easily tested a posteriori by comparing the predicted HI mass distributions of the objects in the sample with the observed ones.

The normalization of the HIMF has been computed by using the first of the approximations to the unbiased minimum variance estimator of the mean number density of galaxies given by Davis & Huchra (1982)

$$h^{-3}n_1 = (h^3 V)^{-1}\left[\sum_i \frac{1}{\Psi(hr_i)} \pm \sqrt{\sum_i \frac{1}{\Psi(hr_i)^2}}\right] \text{Mpc}^{-3} , \qquad (13)$$

where the summation is over all galaxies included in the volume $h^3 V = \omega[(hr_{\max})^3 - (hr_{\min})^3]/3$ limited by $hr_{\min}$ and $hr_{\max}$ in the radial coordinate and by a solid angle $\omega$, and

$$\Psi(hr_i) = \Phi(\mathcal{M}_{\text{HI,lim}}(hr_i))/\Phi(\mathcal{M}_{\text{HI,min}}) , \qquad (14)$$

is the selection function. Equation (13) gives therefore an estimate of the number density of galaxies with HI masses greater than $\mathcal{M}_{\text{HI,min}} \equiv \mathcal{M}_{\text{HI,lim}}(hr_{\min})$. For our sample, $\omega = 0.84\,\text{sr}$ and $hr_{\min} = 20\,\text{Mpc}$, which corresponds to a low-HI-mass cutoff $\mathcal{M}_{\text{HI,min}} = 8.37$. Following Davis & Huchra (1982), we have excluded from the determination of the mean density galaxies at distances larger than $hr_{\max} = 100\,\text{Mpc}$ in order to discard the most uncertain values of $\Psi(hr)$. This restriction however has not been applied in the calculation of the shape of the HIMFs, since we find that the results are quite insensitive to the value adopted for $hr_{\max}$ (see § 3.3).

### 3.2. HI Line Completeness

In spite of the fact that the galaxies in the field sample have been selected from a (nearly) complete optical magnitude-limited data set, completeness in the HI line is not at all guaranteed. This point is illustrated in Figure 3, in which we have plotted the corrected optical magnitude $m_o$ as a function of the quantity $-2.5\log F_{\text{HI}}$ for all the galaxies in this sample. This diagram shows the existence of a well-defined correlation between both quantities truncated by the optical magnitude cutoff at $m_{o,\text{lim}} \sim 15.7$. It is readily apparent that, at low HI line fluxes, only the optically brightest objects are included in the field sample. To overcome this problem one may, for instance, reformulate equation (8) to include a parametric incompleteness function (STY). However, we have no a priori knowledge of the functional form of the incompleteness correction to the HI flux. Besides, the clumpiness in redshift space (§ 2.1) of the current sample complicates the modeling of this correction by comparing the observed number counts with those expected for a universe with the same spatial distribution. We therefore prefer to restrict the data set at a given HI-flux limit above which the HI line completeness is secure.

In order to estimate the HI line completeness limit for the field sample, we apply a density-independent procedure analogous to the $V/V_{\max}$ test (Schmidt 1968) commonly used in the verification of the homogeneity of the space distribution of objects in flux-limited samples. Our procedure relies on the uniformity of the frequency distributions of the integral conditional probability $P(\mathcal{M}_{\text{HI}}|hr) = \Phi(\mathcal{M}_{\text{HI}})/\Phi(\mathcal{M}_{\text{HI,lim}}(hr))$ for subsets of the field sample with different HI-flux limits. Given a parametrization of the HIMF that is a good representation of the data, $P(\mathcal{M}_{\text{HI}}|hr)$ should be randomly distributed between 0 and 1 if the subsets are complete. The HI line completeness limit can be then identified as the minimum limiting value of the corrected (integrated) HI flux below which the computed distributions of this probability start to show significant departures from uniformity.

The frequency distributions of $P(\mathcal{M}_{\rm HI}|hr)$ for five subsets of the field sample corresponding to five different HI-flux limits, constructed from trial calculations of their respective optimized gaussian HIMFs are shown in Figure 4. No apparent deviations from uniformity are seen for the distributions corresponding to the subsets with the highest HI-flux cutoffs, implying that a gaussian HIMF is a good description of the data (see also the next section). However, as the flux limit is lowered, the counts start to show a noticeable tendency to decrease as $P(\mathcal{M}_{\rm HI}|hr) \to 1$. This is just the behavior expected for flux-limited samples that are incomplete at the faint end. A one-sided Kolmogorov-Smirnov test used to quantify the visual impression detects statistically significant departures from uniformity (i.e., the probability $P_{\rm KS}$ of the null hypothesis is 0.1 or less) for limiting HI fluxes below $\sim 10^{0.4}\,{\rm Jy\,km\,s^{-1}}$. Similar results are obtained for the Schechter model. Since, in addition, the best-fitting parameters of the HIMF for any subset with a HI-flux limit above that value are practically invariant, we take $F_{\rm HI,lim} = 10^{0.4}\,{\rm Jy\,km\,s^{-1}}$.

Before discussing in detail the results of the analytic fits to the HIMFs, it is relevant to underscore that the 532 galaxies included in the complete HI-flux-limited subsample (i.e., those with $F_{\rm HI} \geq 10^{0.4}\,{\rm Jy\,km\,s^{-1}}$): i) have $hr \lesssim 150\,{\rm Mpc}$ so they provide a valid reference for the comparison of the HI properties of galaxies in a substantial portion of our local universe; ii) are all detected in HI, because the estimated upper limits to the HI flux for nondetections are well below the adopted HI-flux cutoff (see Fig. 3); and iii) have angular diameters above $0.\!'5$ so the objects with the most uncertain apparent sizes are automatically excluded from the determination of the HIMFs (see also Appendix A).

### 3.3. Results

#### 3.3.1. Gaussian model

In Table 1, we list the best estimates of the normalization and shape parameters (cols. [3]–[5]) of the gaussian HIMF (eq. [10]) for different morphological subsets of the complete HI line subsample, along with the number of galaxies used $N$ (col. [2]). The quoted uncertainties have been calculated from the $1\sigma$ error contours (68.2% confidence intervals) drawn in the $(\langle \mathcal{M}_{\rm HI} \rangle, \sigma_{\mathcal{M}_{\rm HI}})$ plane in Figure 5. It is clear from this figure that the uncertainties in the two parameters are highly correlated in the sense that a lower mean HI mass leads to a larger dispersion. Table 1 and Figure 5 show that, within the errors, the gaussian HIMFs for the different spiral types have identical dispersions. The values of the mean HI masses are relatively more spread, the lowest value corresponding to the Sa-Sab group and the highest to the Sb galaxies. The differences among the inferred theoretical distributions for the various morphological subgroups are statistically insignificant, except for the Sb galaxies. As we discuss below, the high mean HI mass found for the Sb-type systems reflects a true deficiency of low HI-mass objects rather than an excess of Sb galaxies with very large HI contents. HG84 found also identical morphological dependences although affected by larger uncertainties. The slightly higher $\mathcal{M}_{\rm HI}$ means quoted in this previous study are consistent with the fact that they were inferred from direct fits to the observed HI mass distributions and no Malmquist bias correction was applied. To counteract this bias, Roberts & Haynes (1994) used in their investigation of the morphological dependence of integral galaxy properties, an approximately volume-limited sample constructed with members of the Local Supercluster. In spite of the fact that no other selection effects were investigated, the values they provide for $\langle \mathcal{M}_{\rm HI} \rangle$ show, once the difference in the adopted cosmological scales is taken into account, a good agreement with ours within the errors.

In order to test for the goodness-of-fit of the optimized gaussian models, we have compared the observed HI mass distributions of different subsets of the complete HI-flux-limited sample with the expected ones, calculated from the expression

$$p(\mathcal{M}_{\rm HI}|hr_1,\ldots,hr_N) = \sum_{i=1}^{N} p(\mathcal{M}_{\rm HI}|hr_i) \ . \tag{15}$$

The observed HI mass histograms for different morphological subgroups are shown in Figure 6. Superposed dotted curves are plots of equation (15) using the best-fitting parameters in Table 1 for each morphological class. Estimates of the goodness of the fits, given by the $\chi^2/\nu$ statistic, are calculated by assuming poissonian errors and using only bins with four or more galaxies. The number of degrees of freedom $\nu$ is taken equal to the number of such bins, minus one, in order to account for the normalization implicit in equation (15). It is readily apparent from Figure 6 that the gaussian model provides an excellent fit to the observed HI mass histograms of all types of giant spirals.

Similarly, the independence of the HIMF on position has been checked by considering HI mass distributions in different distance intervals. Figure 7 shows a breakup of the complete HI line subsample into two distance bins: $hr \leq 60$ and $hr > 60\,{\rm Mpc}$. The dotted lines are plots of equation (15) using the first set of parameters in Table 1. The reduced chi-square values for the two distance subsets show no significant deviations. This result implies that the HI contents, and morphological types, of the galaxies that lie in the periphery of the main ridge of the Pisces-Perseus supercluster, which dominate the counts in the nearest distance bin, are similar to those of the more distant objects. Hence, the adoption of the hypothesis of a HIMF with a universal form in § 3.1 is fully justified for our data. Finally, the reliability of the adopted HI line completeness limit has also been tested by examining the frequency distributions with HI mass in subsets of the field sample affected by stricter HI-flux limitations. The two panels in Figure 8 show the HI-mass histograms for two subsamples containing all the objects with HI fluxes above $10^{0.8}$ (upper) and $10^{0.6}$ Jy km s$^{-1}$ (lower). The gaussian model corresponding to $F_{\rm HI,lim} = 10^{0.4}$ Jy km s$^{-1}$ (dotted lines) reproduces remarkably well the two observed HI mass distributions. This result supports therefore our decission of adopting this later value of the HI flux as the completeness limit for the field giant spiral sample.

### 3.3.2. Schechter model

The parameters of the Schechter model are listed in columns (6)–(8) of Table 1, whilst Figure 9 shows the 1$\sigma$ error ellipses in the $(\mathcal{M}^*_{\rm HI}, \alpha_{\rm HI})$ plane (notice that the two adjustable parameters of the Schechter model are also highly correlated). As in the gaussian parametrization, the Schechter HIMFs are very similar for the different giant spiral types, except for the Sb galaxies, which have a significantly smaller value of the low-mass-end slope $\alpha_{\rm HI}$. In all cases, the optimized values of this parameter are greater than $-1$, implying a decline in the number of low-HI-mass objects relative to the more massive ones. However, due to the small size of the morphological subsets, error ellipses are large and permit divergent HIMFs at the faint end for all but the Sb galaxies. For this latter morphological type, the statistically acceptable values of $\alpha_{\rm HI}$ imply, in agreement with the gaussian model, that the HIMF is sharply depressed at the low-mass end. As, on the other hand, the values of $\mathcal{M}^*_{\rm HI}$ are fully compatible with the rest, we conclude that the discrepancy shown by the HIMF of the Sb-type systems is exclusively due to the absence of objects with low HI mass. This interpretation is supported by reports of a scarcity of intrinsically faint Sb galaxies with narrow HI line linewidths (e.g., Roberts 1978; Sandage, Binggeli, & Tammann 1985).

Recently, Briggs & Rao (1993) directly fitted a Schechter model to the observed HI mass distribution of a volume-limited (systemic velocity less than 1000 km s$^{-1}$) subset of the Fisher & Tully (1981) HI catalog of galaxies in the Local Supercluster. In spite of the lack of galaxies with HI masses above $10^{10}\,h^{-2}\,M_\odot$ in their restricted sample, the range of best-fitting values for $\mathcal{M}^*_{\rm HI}$ reported by these authors (9.25–9.55) agrees remarkably well with our estimations. In contrast, the faint-end slopes of their best fits correspond to HIMFs that are roughly flat or slightly raising toward low masses. The discrepancy between our values of $\alpha_{\rm HI}$ and those obtained by Briggs & Rao (1993) may be explained by the fact that their data set contained a large number of dwarf irregular systems, excluded from our sample. Yet it must be also taken into account that the scarcity of galaxies with low HI mass in our sample is expected to introduce a large uncertainty in the determination of the shape of the HIMF at the low-mass end. Indeed, tests of the goodness-of-fit of the Schechter model reveal that this parametrization describes our data as well as the gaussian one. This is illustrated in Figure 10, where we have plotted the HI mass distributions for all the galaxies in the complete HI line sample predicted by these two functional forms, along with the observed histogram. In the insert in the upper-left corner of this figure, we have drawn schematic representations of the *normalized* gaussian and Schechter HIMF, using our estimates of the mean number density of galaxies and optimized model parameters given in Table 1. It can be seen that the two theoretical distributions follow closely the observed one. The similarity of the predicted HI-mass distributions, in spite of the discrepancy of the model HIMFs at the low-HI-mass end, is due to the high HI-flux cutoff adopted, which only allows to sample the 'brightest' portion of the HIMF, where the discrepancies between the two optimized models are small. Accordingly, it is reasonable to expect that any other function mimicking the form of the two models investigated here for $\mathcal{M}_{HI} \gtrsim 9$ would also give reasonable fits to our data, regardless of its behavior at the low-mass end.

## 4. The Linear Dependence of $\mathcal{M}_{HI}$ on $\mathcal{D}_o$

In a comparative study of the galaxy properties commonly used in the measurement of the HI content, HG84 found that, for any spiral type, $\mathcal{M}_{HI}$ and $\mathcal{D}_o$ followed a tight linear relation, which made $\mathcal{D}_o$ an excellent diagnostic tool for the HI mass. Provided that the HI-mass measurements are normally distributed around their expectation values with a scatter $\sigma_{\mathcal{M}_{HI}|\mathcal{D}_o}$ independent of galaxy size, unbiased estimates of the coefficients of this relationship can be derived by using in the minimization of equation (9) a straightforward generalization of the gaussian model (eq.[10])

$$\phi(\mathcal{M}_{HI})d\mathcal{M}_{HI} \propto \exp\{-0.5[(\mathcal{M}_{HI} - a_{HI} - b_{HI}\mathcal{D}_o)/\sigma_{\mathcal{M}_{HI}|\mathcal{D}_o}]^2\}d\mathcal{M}_{HI} , \qquad (16)$$

where $p(\mathcal{M}_{HI,i}|hr_i)$ is now substituted by $p(\mathcal{M}_{HI,i}|hr_i, \mathcal{D}_{o,i})$, i.e., the conditional probability that a galaxy in a HI-flux-limited sample at a distance $hr_i$ *and with an optical diameter $\mathcal{D}_{o,i}$* has the observed HI mass $\mathcal{M}_{HI,i}$. Notice that this is just a linear least-squares fitting to the $\mathcal{M}_{HI}$-$\mathcal{D}_o$ relation that takes into account the flux limitation of our sample (i.e., corrected for Malmquist bias). The optimized values of the three adjustable parameters ($a_{HI}$, $b_{HI}$, and $\sigma_{\mathcal{M}_{HI}|\mathcal{D}_o}$) are given in Table 2. Their quoted uncertainties have been calculated from the $1\sigma$ confidence intervals (eq.[12]) corresponding to a $\chi^2$ distribution with 3 degrees of freedom.

The dispersions about the regression lines of the different morphological subgroups are similar, although they show a slight tendency to increase toward earlier types. We estimate that the uncertainty in the values of $\mathcal{M}_{HI}$, due to measurement errors in the HI flux integrals, is typically 0.10. Thus, most of the observed scatter is real and reflects variations in the intrinsic galaxy properties that are not accounted for by the simple linear model adopted here. In this one respect, it is worthwhile to comment that when $\mathcal{D}_o$ is replaced by $\log(h^2 L_o)$ in equation (16), the corresponding dispersions are about 25% larger. This result implies that, when only linear dependences are assumed, the linear optical diameter of a galaxy gives a more accurate estimate of HI mass than does the optical luminosity, in agreement with the results of HG84.

The two coefficients describing the regression line, $a_{\rm HI}$ and $b_{\rm HI}$, are nearly identical from Sa to Sbc. For these types, the statistically acceptable values of $b_{\rm HI}$, which is highly correlated with $a_{\rm HI}$, inferred from the $1\sigma$ error contours of the two regression coefficients (not shown here), range between 0.9 and 1.5. Only the optimized slope for the Sc galaxies, with statistically acceptable values within the range 1.6–1.9, supports the near constancy of $\overline{\Sigma}'_{\rm HI}$ and, provided that the ratio of optical to HI diameters is invariant, that of the *true* global HI surface density (calculated averaging the total HI mass over the HI disk), claimed to hold for the *entire* spiral population by Shostak (1978), Hewitt, Haynes, & Giovanelli (1983), Giovanelli & Haynes (1983), and HG84. The use in previous investigations of small, especially selected, samples may explain this result. We note, for instance, that with the exception of HG84, who explicitly investigated the morphological type dependence of the $\mathcal{M}_{\rm HI}$-$\mathcal{D}_o$ relationship, the steeper slopes found in these early studies are consistent with the fact that they were inferred from data sets largely dominated by Sc galaxies. As for the implications of the optimized values of $b_{\rm HI}$ obtained here on the possible invariance of the true global HI surface density, it should be emphasized that the optical and HI sizes may not be directly proportional, since the optical light and the HI emission do not arise from the same locations. In particular, the HI distribution of giant spirals is typically more extended than is the light and often shows a strong central depression which becomes more conspicuous towards earlier types (see e.g., Haynes & Broeils 1995). Such depressions are sometimes larger than the optical disks themselves as, for example, in the case of M31 (Brinks & Bajaja 1986). While it may be the only available measure of disk size, the optical diameter clearly does not well represent the extent of the HI disk in such extreme cases. At any rate, our results in the form of both the large values of the slope and the substantial reduction in the scatter (compare the values of $\sigma_{\mathcal{M}_{\rm HI}|\mathcal{D}_o}$ in Table 2 with those of $\sigma_{\mathcal{M}_{\rm HI}}$ in Table 1) do confirm that the optical diameter by itself is indeed a much more important parameter in predicting the HI mass than is the morphological type.

The correlation between $M_{\rm HI}$ and $D_o$ plays a fundamental role in the measure of the environment-driven depletion of the HI content. It is common practice to quantify the HI deficiency by means of a parameter $DEF$ equal to the difference, in logarithmic units, between the observed HI mass $M_{{\rm HI},i}$ and that expected for a galaxy with the same morphological type $T_i$ and linear optical diameter $D_{o,i}$ unaffected by external influences, that is

$$DEF = \widehat{\mathcal{M}}_{\rm HI}(T_i, D_{o,i}) - \mathcal{M}_{\rm HI,i} \, . \tag{17}$$

The expectation value of the (logarithm of the) HI mass, $\widehat{\mathcal{M}}_{\rm HI}(T, D_o)$, can be empirically derived from the linear regression of $\mathcal{M}_{\rm HI}$ on $\mathcal{D}_o$ inferred from a reference galaxy sample such as the one described here (i.e., using the values of Table 2). The assumed near constancy of the global hybrid HI surface density for all the giant spiral types has suggested the use of an approximation to equation (17) as the difference between the expected and observed $\log(M_{\rm HI}/D_o^2)(T) \propto \overline{\Sigma}'_{\rm HI}(T)$. This relation has the appeal of being distance independent. The current results however imply that $\overline{\Sigma}'_{\rm HI}$ has a relatively strong built-in dependence on $\mathcal{D}_o$ for types earlier than Sc. Since the galaxy population in dense environments is biased against late-type spirals, the investigations of the HI deficiency in galaxy clusters carried out so far are likely to have overestimated (underestimated) the gas deficiency of the largest (smallest) objects.

## 5. Summary and Conclusions

A complete HI-flux-limited sample consisting of 532 field galaxies of types Sa-Sc has been used to determine their HIMF. This sample has been drawn from a larger data set of 934 CGCG objects observed in the HI line which lie in the lowest density environments of the Pisces-Perseus supercluster region. The shape of the HIMFs has been determined using a standard parametric maximum-likelihood technique independent of density inhomogeneities. Both a gaussian and a Schechter parametrization of the form of the HIMF have been tested, with the following results:

(1) Our best estimates of the adjustable parameters of the gaussian HIMF model (mean HI mass and dispersion) show little variation among the different morphological types of giant spirals, with the exception of the Sb galaxies which exhibit a significantly larger mean HI mass. These estimates have been shown to be uncorrelated with the location of the galaxies and robust vis-à-vis changes in the adopted HI completeness limit.

(2) Similarly, the best estimates of the free parameters of the Schechter model (characteristic HI mass and low-mass-end slope) are, in general, independent of morphology. Again the Sb galaxies are the exception, with a HIMF that has a significantly steeper decrease toward the low-mass end. The differences in the HIMF of the Sb-type systems are interpreted in terms of a scarcity of small objects.

(3) Both the gaussian and Schechter HIMFs provide good fits to our data in spite of their different behavior at the low-HI-mass end. This is a consequence of the fact that the low-HI-mass portion of the HIMF, where the two optimized models show the largest discrepancy, is largely excluded from our complete HI-flux-limited sample due to the high HI-flux cutoff adopted.

The maximum-likelihood technique has also been applied to infer Malmquist bias-free scaling laws between the HI mass and the linear optical diameter for the different types of giant spirals. The tightness of the correlations indicate that the HI content is far more accurately predicted by the disk size than by the morphological type alone, in good agreement with the findings of earlier investigations. Yet the inferred relationships do not support the traditionally advocated near constancy of the $M_{\rm HI}/D_{\rm o}^2$ ratio for the entire spiral population, but imply a significant decrease of this ratio with increasing disk size for all but the Sc galaxies. The implications of this result on the invariance of the true global HI surface density are however not straightforward, as they depend on the poorly known relationship between the optical and the HI sizes.

Both the HIMFs and the $\mathcal{M}_{\rm HI}$-$\mathcal{D}_{\rm o}$ linear relationships inferred in the present study will be used in a forthcoming paper to reexamine the HI deficiency patterns of the giant spiral population in a large sample of galaxy clusters, and to investigate the origin of the gas deficiency.

JMS acknowledges support by the United States-Spanish Joint Committee for Cultural and Educational Cooperation and the Dirección General de Investigación Científica y Técnica through Postdoctoral Research Fellowships. This work was supported by grants AST-9115459 to RG, and AST-9023450 and AST-9218038 to MPH.

## A.   Internal Absorption Correction to Apparent Optical Magnitude for Giant Spirals

Following Holmberg (1958) it is possible to derive the amount of internal absorption in galaxies at any wavelength from the observed dependence of their corresponding global surface magnitude on inclination (axial ratio), provided that the apparent sizes are not significantly affected by inclination. Because visual diameters are roughly proportional to face-on values (e.g., Burstein, Haynes, & Faber 1991; Chołoniewski 1991; Huizinga & van Albada 1992; Giovanelli et al. 1994), the application of Holmberg's method to the AGC galaxies, which have visually determined sizes, is fully justified.

We have followed the same general procedure described in the Appendix A of HG84. The current sample however is based on all the CGCG giant spirals included in the AGC instead of on the UGC members. In order to reduce the impact of observational uncertainties in the inferred relations, only galaxies with galactic extinction corrections less than 1 mag and with $1\rlap{.}'3 \leq a \leq 4\rlap{.}'0$ have been considered. The upper limit on $a$ is set to eliminate those galaxies with the most uncertain Zwicky apparent magnitudes (due to their large visual sizes and the way in which these magnitudes were estimated). The lower boundary is introduced to avoid the artificial coupling between the values of $b/a$ and $a$ arising from the existence of a lower limit of $\sim 0\rlap{.}'1$ for the angular dimensions of the objects that can be unambiguously measured in the POSS prints. As a result, for any given value of $a$, there are very few galaxies with axial ratios below $0\rlap{.}'1/a$. The existence of this resolution limit (a similar effect due to sky seeing for a sample of galaxies with isophotal radii is discussed in Giovanelli et al. 1994) is illustrated in Figure 11, which shows the medians of the observed axial ratios $b/a$ and their associated upper and lower quartiles computed over bins of $a$ for *all* the galaxies listed in the AGC. It is clear from this figure that the values of the axial ratio for highly inclined galaxies are more likely to be overestimated as the apparent size decreases, which introduces the observed ficticious dependence between the values of $b/a$ and $a$. It should be pointed out however that this result is not in contradiction with the claimed independence of visual diameters on inclination, because catalogs such as the UGC are diameter-limited ($a_{\rm lim} \sim 1'$) and, hence, relatively insensitive to this effect.

Figure 12 shows plots of the dependence of the global optical surface magnitude, $\overline{\Sigma}_{\rm o}^{-i} = m^{-i} + 5\log a$ corrected to face-on viewing without any correction for internal absorption (see § 2.2.4) for the different morphological classes. Local averages of $\overline{\Sigma}_{\rm o}^{-i}$ have been computed in bins of $\log(b/a)$ to increase the weight of the systems with the highest inclinations. In the calculation of the linear regressions only those averages obtained with more than ten galaxies have been considered (filled symbols). In agreement with the results of HG84, Sb and Sc galaxies exhibit the largest internal extinctions, which decrease moderately toward earlier types. The inferred amount of absorption for the earliest disk systems should be regarded with caution, however, because the presence of a substantial bulge can affect the measure of the minor diameter, shifting $b/a$ toward artificially large values.

For a given morphological type, the correction to apparent magnitude for internal absorption $\Delta m^i$ is simply given by

$$\Delta m^i = -k \log(b/a) , \qquad (A1)$$

where $k$ is the slope of the corresponding linear regression listed in Table 3.

# FIGURE CAPTIONS

Fig. 1.— *Top panel:* Sky distribution of all CGCG galaxies in the region of the Pisces-Perseus supercluster bounded by $22^{\rm h} \leq$ R.A. $\leq 3^{\rm h}$, $0° \leq$ Dec. $\leq 40°$. *Middle panel:* CGCG galaxies in that region with local surface density equal to or above $3\,\text{deg}^{-2}$ (see text). The locations of the most prominent galaxy clusters delineating the supercluster are indicated. *Bottom panel:* Galaxies with local surface density below that limit. The number $N$ of galaxies included in each subset is quoted on top of each panel.

Fig. 2.— Velocity distributions for the galaxies with known redshift in the corresponding panels in Figure 1. The superposed smooth curves represent the expected velocity distributions for a spatially homogeneous population of objects with the same apparent magnitude distribution as the observed samples. The dotted histogram in the lower panel shows the velocity distribution for the galaxies with local surface densities below $1\,\text{deg}^{-2}$. Bins in velocity are $250\,\text{km}\,\text{s}^{-1}$ wide for the solid histograms and $500\,\text{km}\,\text{s}^{-1}$ wide for the dotted one.

Fig. 3.— Corrected optical magnitude plotted against observed HI flux (in magnitude units), for all the objects in the field giant spiral sample. Arrows identify nondetections plotted at their estimated upper limits (see text).

Fig. 4.— Frequency distribution of the probability $P(\mathcal{M}_{\rm HI}|hr) = \Phi(\mathcal{M}_{\rm HI})/\Phi(\mathcal{M}_{\rm HI,lim}(hr))$ for galaxies in the field sample with HI fluxes above different limits $S_{\rm HI,lim}$. The probabilities for each of the HI-flux-limited subsets are calculated using their respective optimized gaussian HIMF. The binning has been varied in order to mantain a similar number of objects per probability interval for a uniform distribution. Mean values and poissonian error bars have been normalized to unity. $P_{\rm KS}$ gives the probability that the observed distributions are uniform. $N$ gives the number of objects included in each subset.

Fig. 5.— $1\sigma$ error contours for the free parameters of the gaussian HIMFs inferred from the complete HI line subsample. Error ellipses for the entire data set (thick solid line), as well as for different morphological subgroups (thin curves), are plotted. The thick dashed line is the $2\sigma$ error contour for all spiral types. Crosses mark the position of the maximum-likelihood solutions.

Fig. 6.— Observed distribution of HI mass for the galaxies in the complete HI line subsample according to their morphological type. Dotted curves represent the distributions predicted by the optimized gaussian models corresponding to the parameters listed in Table 1. The values of the reduced chi-squares have been calculated using errors given by Poisson statistics. The total number $N$ of galaxies included in each subsample is also quoted.

Fig. 7.— Same as in Fig. 6, but breaking the complete HI line subsample into two different distance ranges, nearer than and beyond $60\,h^{-1}$ Mpc.

Fig. 8.— Same as in Fig. 6, but for all the objects included into two subsets of the complete HI line subsample generated by applying stricter HI-flux cutoffs. The theoretical curves are calculated using the best-fitting parameters determined for the adopted completeness limit $F_{\rm HI,lim} = 10^{0.4}\,{\rm Jy\,km\,s^{-1}}$.

Fig. 9.— Same as in Fig. 5, but for the Schechter model.

Fig. 10.— Distributions of HI mass for all the objects in the complete HI line subsample predicted by the gaussian (solid curve) and Schechter (dot-long dashed curve) models. The observed histogram is also plotted. Schematic representations of the normalized HIMFs corresponding to both parametrizations are drawn in the insert in the upper-left-corner. The dotted section of the theoretical curves indicate extrapolations below the sample's low-HI-mass cutoff $\mathcal{M}_{\rm HI,min} = 8.37$.

Fig. 11.— Local medians of the axial ratio, plotted against apparent diameter, for all the galaxies in the AGC. Error bars are upper and lower quartiles. The solid curve identifies the lower limit for $b/a$ resulting from the constraint $b \geq 0\farcs1$.

Fig. 12.— Variation of the global optical surface magnitude, uncorrected for internal extinction, with axial ratio for different types of spirals. Open symbols represent averages with fewer than 10 galaxies not included in the fits. The total number of galaxies in each bin is given along with the corresponding $1\sigma$ error bars.